\documentclass[twocolumn,superscriptaddress,preprintnumbers,notitlepage,amsmath,10pt,aps,pra]{revtex4-1}
\usepackage{graphicx}
\usepackage{color}
\usepackage{subfigure}

\def\change#1{\textcolor{blue}{#1}}
\usepackage{makeidx}
\makeindex

\begin{document}
\title{Quantum measurements of atoms using cavity QED} 

\affiliation{Department of Physics, Heriot-Watt University, Edinburgh EH14 4AS, UK}
\affiliation{School of Physics and Astronomy, University of Leeds, Leeds  LS2 9JT, UK} 
\affiliation{National Institute of Informatics, 2-1-2 Hitotsubashi, Chiyoda-ku, Tokyo 101-8430, Japan
}

\author{Adetunmise C. Dada}
\email{acd8@hw.ac.uk}
\affiliation{Department of Physics, Heriot-Watt University, Edinburgh EH14 4AS, UK}
\author{Erika Andersson}
\affiliation{Department of Physics, Heriot-Watt University, Edinburgh EH14 4AS, UK}
\author{Martin L. Jones}
\affiliation{School of Physics and Astronomy, University of Leeds, Leeds  LS2 9JT, UK}
\author{Vivien M. Kendon}
\affiliation{School of Physics and Astronomy, University of Leeds, Leeds  LS2 9JT, UK}
\author{Mark S. Everitt}
\affiliation{National Institute of Informatics, 2-1-2 Hitotsubashi, Chiyoda-ku, Tokyo 101-8430, Japan
}
\affiliation{School of Physics and Astronomy, University of Leeds, Leeds  LS2 9JT, UK}

\begin{abstract}
Generalized quantum measurements are an important extension of projective or von Neumann measurements, in that they can be used to describe any measurement that can be implemented on a quantum system.
We describe how to realize two non-standard quantum measurements using cavity quantum electrodynamics (QED). The first measurement optimally and unabmiguously distinguishes between two non-orthogonal quantum states. The second example is a measurement that demonstrates superadditive quantum coding gain. The experimental tools used are single-atom unitary operations effected by Ramsey pulses and two-atom Tavis-Cummings interactions. 
We show how the superadditive quantum coding gain is affected by errors in the field-ionisation detection of atoms, and that even with rather high levels of experimental imperfections, a reasonable amount of superadditivity can still be seen. 
To date, these types of measurement have only been realized on photons. It would be of great interest to have realizations using other physical systems. This is for fundamental reasons, but also since quantum coding gain in general increases with code word length, and a realization using atoms could be more easily scaled than existing realizations using photons.
\end{abstract}
\maketitle

\section{Introduction}
Generalized quantum measurements or probability operator measures (POMs), also called positive operator valued measures (POVMs), are important mathematical tools for quantum communication and quantum information processing~\cite{helstrom}. 
They are naturally able to describe imperfections and errors in real experimental measurements. In addition, there are also situations where it is advantageous to deliberately engineer a measurement that is not a projective measurement. This is frequently the case when distinguishing between quantum states~\cite{helstrom, chefles2000}. The simplest such example is when distinguishing between two non-orthogonal states without error~\cite{Ivanovic1987257,Dieks1988303,Peres198819}. 
In addition, knowledge of optimal measurement strategies may be useful in placing tight bounds on other quantum operations such as quantum cloning~\cite{chefles2000, PhysRevA.73.062319}.

In this paper, we describe how to realize two examples of non-standard quantum measurements using the tools of cavity QED. The methods we describe could, however, be applied also more generally for realizing other generalized quantum measurements.
The first measurement is optimal unambiguous discrimination of non-orthogonal quantum states, also known as the Ivanovic-Dieks-Peres (IDP) measurement. This task is relevant for quantum information and communication systems as well as for quantum key distribution (QKD)~\cite{RevModPhys.74.145}. The IDP measurement is optimal  for the B92 QKD protocol~\cite{PhysRevLett.68.3121}, although this was not immediately recognised. 
To date, all the realizations of the IDP measurement have been optical~\cite{PhysRevA.54.3783, PhysRevA.63.040305,PhysRevLett.93.200403}. Nevertheless, generalized quantum measurements could be realized also on ions or atoms using existing experimental techniques \cite{PhysRevA.63.052301, PhysRevA.64.032303, PhysRevA.66.012103}, or using nuclear magnetic resonance~\cite{PhysRevA.71.042307}.

The second example is the measurement required to demonstrate that quantum channel capacities can be superadditive. In this case, at least two uses of a quantum channel, and a collective measurement of the resulting code block, is required. The quantum coding gain in general grows with the length of the code blocks. Superadditivity has so far only been demonstrated using linear optics~\cite{PhysRevA.69.052329}. Quantum source coding for message compression is another type of quantum coding scheme that has been optically demonstrated~\cite{PhysRevLett.91.217902}, using similar techniques as for the optical demonstration of quantum superadditivity.
 In both cases, the two uses of the quantum channel were encoded using the path and polarization degrees of freedom of a single photon and the states are manipulated using basic linear optical elements (polarising beam splitters and waveplates). While this demonstrated the principle of the measurement, extension of the coding to longer code blocks would be impractical due to problems of scalability. Scalability would require effective photon-photon interactions, which are difficult to realize due to prohibitively large overhead costs~\cite{Knill2001Scheme,RevModPhys.79.135}. 
 
 Generally speaking, some generalized measurements are difficult to realize using linear optics. Therefore it is useful to study how to realize such measurements in other physical systems. For the cavity QED demonstration of superadditivity, we use two atoms and encode each usage of the quantum channel in the state of one atom.  This could in principle be scaled to longer codewords using resources which do not scale exponentially, as for the existing optical realisations. Also, other coding schemes, including quantum source coding, or any other realisation of collective quantum measurements, could be realised in a cavity QED setting employing similar methods. We also estimate how experimental imperfections would affect the measurement. Cavity QED techniques have indeed been applied extensively in exploring the quantum dynamics of atoms and photons in cavities and has been used, for example, in preparing entangled states of atoms~\cite{PhysRevLett.79.1}, performing phase gate operations~\cite{PhysRevLett.83.5166}, doing quantum non-demolition measurements of cavity fields~\cite{citeulike:1587714}, and in experimental studies of the process of decoherence in quantum measurements~\cite{PhysRevLett.77.4887}. In addition, there are a number of QED-type systems in which a cavity QED based scheme can be easily implemented. These include circuit and photonic crystal based systems~\cite{0953-2048-20-8-016, 4868850}.  
\section{Distinguishing between two non-orthogonal states}

Generalized quantum measurements are extensions of projective or von Neumann measurements. Just as for projective quantum measurements, probabilities $p(j)$ for measurement outcomes are calculated using the trace rule 
\begin{equation}
\label{eqn:tracerule}
p(j) = {\rm Tr}(\hat{\rho}\hat{\Pi}_j),
\end{equation} 
where $\rho$ is the measured state and $\hat{\Pi}_j$ is the measurement operator corresponding to outcome $j$. The fact that probabilities are positive means that all eigenvalues of the $\hat{\Pi}_j$ are positive, which is written $\hat{\Pi}_j > 0$, and consequently also that the $\hat{\Pi}_j$ are Hermitian. Also, since the probabilities for all possible outcomes should sum to 1, it follows that 
\begin{equation}
\label{eqn:complmeas}
\sum_j\hat{\Pi}_j =  \hat{{\bf I}},
\end{equation} 
where $\hat{{\bf I}}$ is the identity operator. What sets generalized quantum measurements apart from projective measurements is that the measurement operators do not have to be projectors. Also, we can have more measurement outcomes than there are dimensions in the measured quantum system. 

The Ivanovic-Dieks-Peres (IDP) measurement~\cite{Ivanovic1987257,Dieks1988303,Peres198819} is a generalized measurement that distinguishes between two non-orthogonal states without error, in other words, unambiguously. For the measurement to be error-free, however, one must accept that it will sometimes be inconclusive. The IDP measurement is optimal in the sense that it minimises the probability of an inconclusive measurement outcome.
Suppose that we wish to distinguish without error between two non-orthogonal quantum states 
\begin{eqnarray}
\label{eq:stts}
|\psi_1\rangle = \cos \theta | 1 \rangle -  \sin \theta | 2 \rangle\\
|\psi_2\rangle = \cos \theta | 1 \rangle +  \sin \theta | 2 \rangle
\end{eqnarray}
of a single quantum system such as an atom, where $0<\theta<\pi/4$. 
To start with, let us note that the optimal measurement will depend on the probabilities for preparing these states, i.e. the prior probabilities. Let us note also that if we make a projective measurement in the basis  $\{|\psi_1\rangle,|\psi^{\perp}_1\rangle\}$, with $|\psi^{\perp}_1\rangle=\sin \theta | 1 \rangle + \cos \theta | 2 \rangle$, then the outcome $|\psi^{\perp}_1\rangle$ necessarily indicates that the prepared state was $|\psi_2\rangle$. If we obtain $|\psi_1\rangle$, then we cannot be sure which state was prepared, and the outcome is inconclusive. Similarly, if we choose to measure in the basis $\{|\psi_2\rangle,|\psi^{\perp}_2\rangle\}$, then an outcome $|\psi^{\perp}_2\rangle $ indicates that the state was certainly  $|\psi_1\rangle$ and  the outcome  $|\psi_2\rangle$ yields an inconclusive result. If we are restricted to standard von Neumann measurements, this is the best we can do.

This procedure above is however not  always optimal. If the respective probabilities of preparing $|\psi_1\rangle$ and $|\psi_2\rangle$, i.e., the prior probabilities, are similar, then the generalized measurement that gives the lowest possible probability for the inconclusive result has the measurement operators
\begin{equation}
\label{eqn:genidpmeas}
\hat{\Pi}_1=k|\psi^{\perp}_2\rangle\langle\psi^{\perp}_2|,~~\hat{\Pi}_2=k|\psi^{\perp}_1\rangle\langle\psi^{\perp}_1|,~~\hat{\Pi}_?=\hat{{\bf I}}-\hat{\Pi}_1-\hat{\Pi}_2,
\end{equation}
where $k$ is a positive number which is as large as the positivity of $\hat{\Pi}_?$ will allow, that is, $k=1/(1+\langle\psi_1|\psi_2\rangle)$.
The minimum probability for the inconclusive result is then given by $p(?)$=$|\langle \psi_1|\psi_2\rangle|$.
Let us denote the prior probabilities as $p_1$ and $p_2=1-p_1$. It is easy to verify that the generalized measurement is better than the best projective measurement when
\begin{equation}
\label{eqn:priorprange}
1-\sec ^2(2 \theta)+\sec (2 \theta)>p_1>2 \sin ^2(\theta) \sec ^2(2 \theta).
\end{equation}

The measurement described in Eq.~\eqref{eqn:genidpmeas} can be physically realized as a measurement in a higher dimensional Hilbert space in an orthonormal basis. This follows from Naimark's theorem which states that any generalized measurement can be realized in this way~\cite{helstrom}.
We will devise an experimental realization in terms of such a projective measurement in an extended Hilbert space.
The IDP measurement can then be realized using the following steps:
\begin{enumerate}
\item Extend the initial 2D Hilbert space into a 3D space by adding an extra state $|3\rangle$ which is orthogonal to both initial states $|\psi_1\rangle$ and $|\psi_2\rangle$, resulting in an orthonormal basis \{$|1\rangle$, $|2\rangle$, $|3\rangle$\}.
\item Measure in a basis  $\{|\Pi_1\rangle,|\Pi_2\rangle, |\Pi_3\rangle\}$  where $|\Pi_1\rangle \perp |\psi_2\rangle$ and $|\Pi_2\rangle \perp |\psi_1\rangle$. This measurement can be implemented in two steps:
\begin{enumerate}
\item Perform a unitary operation $\hat{U}$ given by
\begin{equation}
\label{eq:uagain}
\hat{U} = |1\rangle \langle\Pi_{1}|+|2\rangle \langle\Pi_{2}|+|3\rangle \langle\Pi_{?}|.
\end{equation}

\item Do a standard projective measurement in the \{$|1\rangle$, $|2\rangle$, $|3\rangle$\} basis. A detection in the states $|1\rangle$ or  $|2\rangle$ would unambiguously indicate that the unknown state was $|\psi_1\rangle$ or $|\psi_2\rangle$ respectively, while a detection result $|3\rangle$ would make the measurement inconclusive. 

The detection probabilities will therefore be
\begin{eqnarray}
\label{eq:poutIDP}
				p(1|1)&=&|\langle \Pi_1|\psi_1\rangle|^2\nonumber\\
				p(2|2)&=&|\langle \Pi_2|\psi_2\rangle|^2\nonumber\\
				p(1|2)&=&p(2|1) =|\langle \Pi_1|\psi_2\rangle|^2=|\langle \Pi_2|\psi_1\rangle|^2=0\nonumber\\ 
				p(?|1)&=&p(?|2)=  |\langle\psi_1|\psi_2\rangle| 
				\label{eq:poutIDPend}.
\end{eqnarray}

\end{enumerate}
\end{enumerate}
Here $p(k|j)$ denotes the probability of obtaining a result $k$ given a state $j$.

Using the basis state vectors $|1\rangle\equiv[1,0,0]^T$, $|2\rangle\equiv[0,1,0]^T$ and $|3\rangle\equiv[0,0,1]^T$, we work out the unitary operation in~\eqref{eq:uagain} for the optimum measurement to be
\begin{equation}
\label{eq:uagain2}
 \hat U = \frac {1} {\sqrt {2}}\left[
\begin{array}{ccc}
 \tan \theta & -1 & -\sqrt{1-\tan ^2\theta} \\
 \tan \theta & 1 & -\sqrt{1-\tan ^2\theta} \\
  \sqrt{2(1-\tan ^2\theta)} & 0 & \sqrt{2} \tan \theta
\end{array}
\right] \end{equation}

\begin{figure}[floatfix]
\centerline{\includegraphics[width=0.45\textwidth]{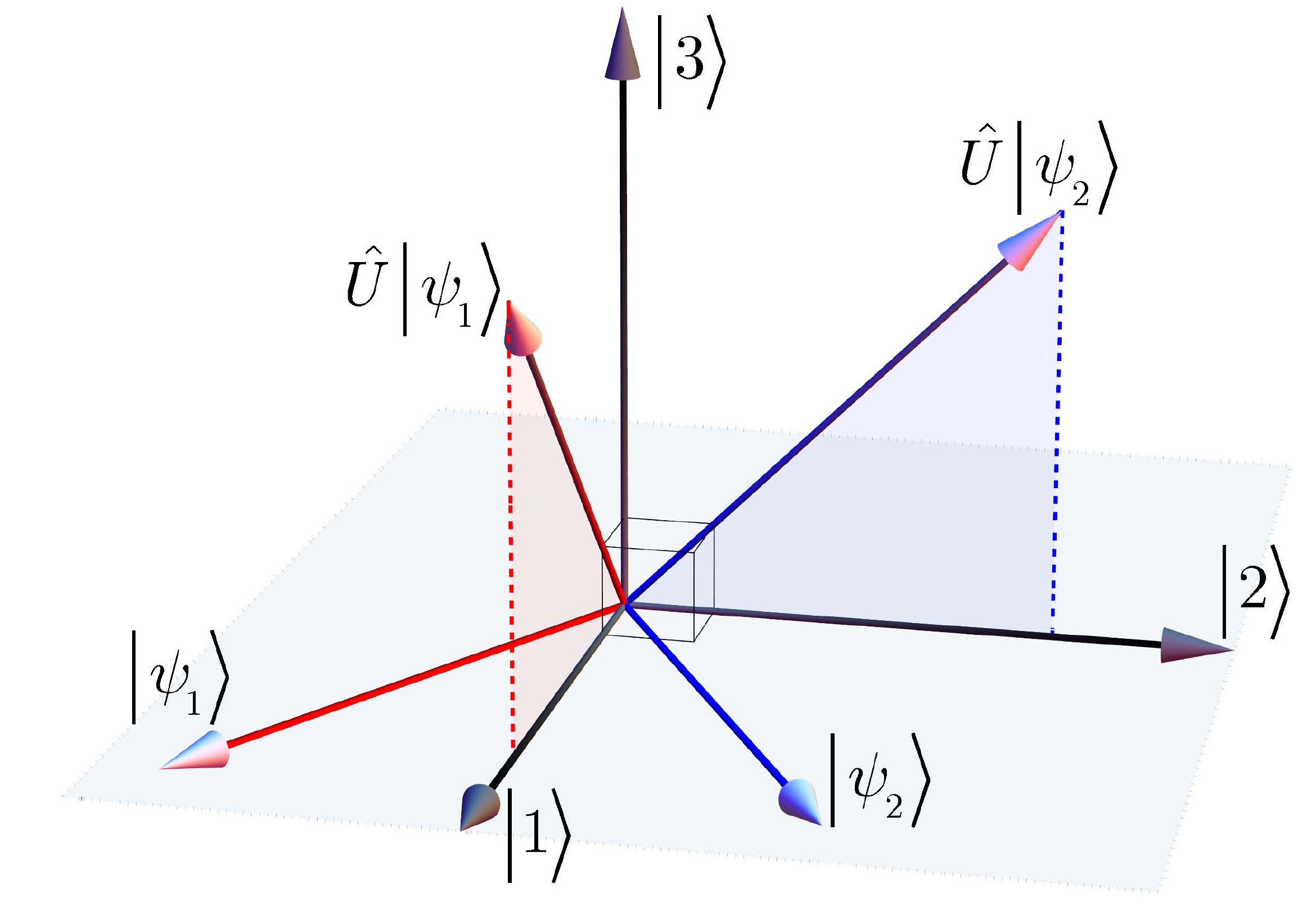}}
\caption{Geometrical representation of the IDP measurement. The initial two-dimensional Hilbert space, where $|\psi_1\rangle$ and $|\psi_2\rangle$ live, is spanned by \{$|1\rangle$, $|2\rangle$\}. This is extended to three dimensions by adding an extra state $|3\rangle$ which is orthogonal to each of the two initial basis states. The measurement consists of a unitary operation $\hat{U}$, followed by a standard projective measurement in the \{$|1\rangle$, $|2\rangle$, $|3\rangle$\} basis. The transformed states $\hat{U}|\psi_1\rangle$ and $\hat{U}|\psi_2\rangle$ are orthogonal to $|2\rangle$ and $|1\rangle$ respectively.}
\label{fig:idp}
\end{figure}

For experimental realization, any unitary $\hat{U}$ may be decomposed into a product of unitary operators coupling two levels at a time~\cite{PhysRevLett.73.58}. Furthermore, especially when there are many outcomes, this decomposition for a generalized quantum measurement may be optimized to use the minimum number of such pairwise operations~\cite{PhysRevA.77.052104}. 
In our case, there is only one extra state, and the realization is straightforward,
\begin{equation}
\label{eq:usequence}
\hat{U}=\hat{T}_{1,2}\hat{T}_{1,3},
\end{equation}
where
\begin{eqnarray}
\hat{T}_{1,2} &=& \left( \begin{array}{ccc}
\frac{1}{\sqrt{2}} & -\frac{1}{\sqrt{2}} & 0 \\
\frac{1}{\sqrt{2}} & \frac{1}{\sqrt{2}} & 0 \\
0 & 0 & 1 \end{array} \right),~{\rm and}\\
\hat{T}_{1,3} &=&\left(
\begin{array}{ccc}
 \tan \theta & 0 & -\sqrt{1-\tan ^2\theta} \\
 0 & 1 & 0 \\
 \sqrt{1-\tan ^2\theta} & 0 & \tan \theta
\end{array}
\right).
\end{eqnarray}
In summary, one performs $\hat{T}_{1,3}$, followed by $\hat{T}_{1,2}$ on the input state $|\psi\rangle \in \{|\psi_1\rangle,|\psi_2\rangle\}$ to obtain $|\psi\rangle'$, followed by a projective measurement of $|\psi\rangle'$ in the basis $\{|1\rangle,|2\rangle,|3\rangle\}$. This will yield an error probability of zero and a minimum probability of an inconclusive result $p(?)=\cos 2\theta$.

\subsection{Cavity QED implementation}
The interaction between an atom and a classical field, resonant or quasi-resonant with the atomic transition between two states $|g\rangle$ and $|e\rangle$, can be used to realize the IDP measurement outlined above. The required unitary operations result from the action of the atom-field Hamiltonian, which is~\cite{RaimondHaroche}

\begin{equation}
\label{eq:hamJC}
\tilde H = \frac{{\hbar \Delta _r }}{2}\sigma _z  - i\hbar \frac{{\Omega _r }}{2}\left[ {e^{ - i\varphi } \sigma _ +   + e^{i\varphi } \sigma _ - } \right].
\end{equation}
Here $\sigma _z=|g\rangle\langle g|-|e\rangle\langle e|$ is the Pauli-Z operator, and $\sigma _ \pm$, the atomic raising and lowering operators, are defined as $\sigma _ +=|g\rangle\langle e|$ and $\sigma _ -=|e\rangle\langle g|$. $\Omega _r$ and  $\Delta _r$ are respectively the classical Rabi frequency and the atom-field detuning, and $\varphi$ is the phase of the classical field with respect to the atomic transition dipole. 

It can be shown from Eq.~\eqref{eq:hamJC} that an interaction lasting for a time $t=\theta/\Omega _r$ with a resonant field having a phase $\varphi$ effects the transformations
\begin{eqnarray}
\label{eq:ramsey2d}
			|g\rangle \longrightarrow &\cos({\theta}/{2})|g\rangle + \sin({\theta}/{2})e^{i\varphi}|e\rangle \nonumber\\
			|e\rangle \longrightarrow &-\sin({\theta}/{2})e^{-i\varphi}|g\rangle + \cos({\theta}/{2})|e\rangle.
\end{eqnarray}
Using the notation $|g\rangle \equiv [1,0]^T$ and $|e\rangle \equiv[0,1]^T$, this corresponds to the operator $\exp (i\tilde{ H}t/\hbar )$ given as
\begin{equation}
\label{eq:ramsey2dm}
R_{g,e}(\theta,\varphi)= \left( \begin{array}{cc}
\cos\frac{\theta}{2} & -e^{-i\varphi}\sin\frac{\theta}{2} \\
e^{i\varphi}\sin\frac{\theta}{2} & \cos\frac{\theta}{2} \end{array} \right) .
\end{equation}%
One may perform the unitary operator $\hat{U}$ for the IDP measurement by setting
\begin{eqnarray}
\label{eq:lalala}
\hat{T}_{1,2} = R_{1,2}({\pi}/{2},0)\\
\hat{T}_{1,3} = R_{1,3}(\vartheta,0),
\end{eqnarray}
where $\vartheta = \cos^{-1}{(\frac{1}{\sqrt{3}})}$, and the $R_{g,e}$ denotes a Ramsey pulse resonant with the transition $|g\rangle\leftrightarrow|e\rangle$.

The physical states representing $| 1\rangle$, $|2\rangle$, and $|3\rangle$ will be chosen based on convenience of experimental realization. One needs to bear in mind, for example, that a direct coupling between $|2\rangle$ and  $|3\rangle$ will not be necessary, and that at the detection stage, outcome $|3\rangle$ represents an inconclusive result. A possible choice of states could be a $^{85}Rb $ ladder of Rydberg states; $|e\rangle\equiv63P$, $|g\rangle\equiv61D$ and $|i\rangle\equiv62P$, which are standard micromaser transitions~\cite{Filipovicz1985}. A final projective measurement which determines the energy level of the atom would be required. This is commonly done by means of field-ionisation detection~\cite{Walther2006,Jones2009a}, which involves passing the Rydberg atoms through an increasing electric field and measuring the energy at which the atom is ionised.

\section{Superadditive measurement}
\label{sec:superad}  
The maximum amount of information which can be reliably transmitted over a given channel is referred to as its capacity. This is generally determined by the information resources (such as code block length and bandwidth) and the noise characteristics of the channel. 
Quantum channels may display superadditivity in classical information capacity~\cite{PhysRevLett.66.1119,PhysRevA.58.146,PhysRevA.61.032309}. This means that
  \begin{equation}
C_n  > nC_1
 \label{eq:sadvtyo},
 \end{equation}
where $C_1$ is the classical information capacity of a single use of the channel, and $C_n$ is the classical information capacity of a combination of $n$ uses of the channel. For classical channels, it holds that
 \begin{equation}
C_n  \change{=} nC_1,
 \label{eq:sadvtyo2}
 \end{equation}
meaning that superadditivity is  displayed only by quantum channels.

This makes it interesting to experimentally demonstrate the superadditivity of quantum channel capacity. In order to do this, it is necessary to carry out quantum coding followed by an appropriate collective quantum measurement. One possible scheme is outlined below.

\subsection{Trine letter states}
Consider a channel coding for sending classical information through a quantum channel with a given
ensemble of quantum states representing the letter states. A clear and simple example of an ensemble which can be used to demonstrate superadditivity in classical capacity of a quantum channel is the qubit trine states.
Suppose we use the set of ternary symmetric states of a qubit, that is $\left\{ {\left| {\psi _0 } \right\rangle ,\left| {\psi _1 } \right\rangle ,\left| {\psi _2 } \right\rangle } \right\}$ known as the \emph{qubit trine states}, with
\begin{eqnarray}
|\psi_0\rangle &=& |0\rangle, \nonumber\\
|\psi _1 \rangle &=& - \frac{1}{2}|0\rangle  - \frac{{\sqrt 3 }}{2}|1\rangle, 
 \nonumber\\
|\psi _2 \rangle  &=& - \frac{1}{2}|0\rangle  + \frac{{\sqrt 3 }}{2}|1\rangle \label{eq:trine3},
\end{eqnarray}
 to transmit information, where \{$|0\rangle$, $|1\rangle$\} is the orthonormal basis set. Using one quantum state drawn from this ensemble we can transmit at most $C_1=0.6454$ bits. This is achieved by sending any two of the states $|\psi_j\rangle$ with probability 1/2 each and distinguishing between these with the optimal measurement~\cite{shor-2002,PhysRevLett.90.167906,PhysRevA.69.052329}.

Using two qubits, there are nine possible  states. It has been shown \cite{PhysRevLett.66.1119} that if only three of these are used, namely
 \begin{eqnarray}
|\psi _{xx} \rangle  &=&|\psi _x \rangle  \otimes |\psi _x \rangle  \nonumber \\
           &=&\frac{1}{2}(1 + \cos \varphi _x )|00\rangle   + \frac{1}{2}\sin \varphi _x \left( {|01\rangle  + |10\rangle } \right) \nonumber \\
           &&+ \frac{1}{2}(1 - \cos \varphi _x )|11\rangle 
 \label{eq:tinecode2},
 \end{eqnarray}
with $\varphi _x  = 2\pi x/3$, where $x=0,~1,~2$, then $I_2=1.3690$ bits of information can be retrieved if the code word states are used with equal probabilities. This is larger than $2C_1(=1.2908)$. The superadditive quantum coding gain (SQCG), per use of the channel, is 
\begin{equation}
\label{eq:sqcg}
(I_2-2C_1)/2=(1.3690-1.2908)/2=0.0391.
 \end{equation}
The measurement used to decode the codewords is the \emph{square-root measurement} with the measurement basis states $\left| {\Pi _{yy} } \right\rangle$ defined as
 \begin{equation}
\left| {\Pi _{yy} } \right\rangle  \equiv \left( {\sum\limits_x {|\psi _{xx} \rangle \langle \psi _{xx} |} } \right)^{ - 1/2} |\psi _{yy} \rangle . 
 \label{eq:sqrtmsre}
 \end{equation}
  
In explicit form the codeword states are
\begin{eqnarray}
|\psi_{00}\rangle &= &|00\rangle, \nonumber\\ 
|\psi _{11} \rangle &=  &[|00\rangle + {\sqrt 3}\left(|01\rangle +|10\rangle\right)+3|00\rangle]/4, \nonumber\\  
|\psi _{22} \rangle  &=  &[|00\rangle - {\sqrt 3}\left(|01\rangle +|10\rangle\right)+3|00\rangle]/4, \label{eq:trine33}
\end{eqnarray}
and the optimal measurement basis is given by
\begin{eqnarray}
|\Pi_{00}\rangle &= &\cos \left(\gamma /2\right)|00\rangle-\sin \left(\gamma /2\right)|11\rangle\nonumber,\\
|\Pi _{11} \rangle &=  &[\sin \left(\gamma /2\right)|00\rangle+\cos \left(\gamma /2\right)|11\rangle]/\sqrt{2}\nonumber\\
&&+(|01\rangle+|10\rangle)/2, \nonumber\\
|\Pi _{22} \rangle &=  &[\sin \left(\gamma /2\right)|00\rangle+\cos \left(\gamma /2\right)|11\rangle]/\sqrt{2}\nonumber\\
&&-(|01\rangle+|10\rangle)/2, \nonumber\\
|A\rangle &= &\left[|01\rangle -|10\rangle\right]]/\sqrt{2},
 \label{eq:mbas14}
\end{eqnarray}
where
\begin{eqnarray}
\cos({\gamma}/{2})&=&(\sqrt{2}+1)/{\sqrt{6}}, {\rm~and}\nonumber\\
\sin({\gamma}/{2})&=&(\sqrt{2}-1)/{\sqrt{6}}.
\end{eqnarray}
The outcome corresponding to the state $|A\rangle$ will never occur, since all codeword states are orthogonal to $|A\rangle$. This state merely completes the 4-dimensional basis.

The states \eqref{eq:mbas14} define an entangled measurement basis, and the implementation of the measurement will require similar resources as a Bell measurement, including entangling interactions.
The Bell states are the maximally entangled states
\begin{eqnarray}
\label{eqn:bells}
|\Phi^\pm \rangle &=& (|00\rangle \pm |11\rangle)/\sqrt{2}\nonumber\\ 
|\Psi^\pm \rangle &=& (|01\rangle \pm |10\rangle)/\sqrt{2},
\end{eqnarray}
and a Bell measurement is a projection in this basis. This can be achieved by first performing a unitary transformation $\hat{U}_B$
\begin{equation}
\label{eqn:uprep1}
\hat{U}_B = \left| {00} \right\rangle \left\langle {\Psi ^ +  } \right| + \left| {01} \right\rangle \left\langle {\Phi ^ +  } \right| + \left| {10} \right\rangle \left\langle {\Psi ^ -  } \right| + \left| {11} \right\rangle \left\langle {\Phi ^ -} \right|
\end{equation}
on the input Bell state, followed by a projective measurement in the \{$|00\rangle$, $|01\rangle$, $|10\rangle$, $|11\rangle$\} basis which we refer to as the computational basis. Any other transformation which takes each of the Bell states respectively to any permutation of the computational basis states, up to global phases, would also do.

The superadditive measurement can be realized in a similar fashion by making a unitary transformation $\hat{U}_{\rm sa}$ on the input states, and following this by a projective measurement in the computational basis. From Eq.~\eqref{eq:mbas14}, $\hat{U}_{\rm sa}$ is given as
\begin{equation}
\label{eq:saddm1}
\hat{U}_{\rm sa}= \left| {00} \right\rangle \left\langle {\Pi_{00}} \right| + \left| {01} \right\rangle \left\langle {\Pi_{11}  } \right| + \left| {10} \right\rangle \left\langle {A } \right| + \left| {11} \right\rangle \left\langle {\Pi_{22}} \right|.
\end{equation}
In matrix notation, it takes the form
\begin{equation}
\label{eq:saddm}
\hat{U}_{\rm sa}=\frac{1}{2}
\left[
\begin{array}{cccc}
 2\cos (\gamma/2) & 0 & 0 & -2\sin (\gamma/2) \\
 \sqrt{2}\sin (\gamma/2)&
   1& 1&
 \sqrt{2}  \cos(\gamma/2)\\
 0 & \sqrt{2} &
   -\sqrt{2} & 0 \\
 \sqrt{2}\sin(\gamma/2)&
   -1 & -1
   & \sqrt{2}\cos (\gamma/2)
\end{array}
\right].
\end{equation}

Superadditivity can in general only be achieved when an appropriate \emph{collective} POM is chosen, namely, detection in an entangled measurement basis \cite{PhysRevLett.90.167906}. 
An SQCG of $0.011\pm 0.003$ has been experimentally demonstrated by the previously mentioned optical implementation~\cite{PhysRevA.58.146}. 

\subsection{Cavity QED realization}

The unitary operation $\hat{U}_{\rm sa}$ needed to realize this measurement 
can be decomposed in terms of single atom operation and entangling interactions. The single atom rotations correspond to Ramsey pulses $\hat{R}_1(\theta,\varphi)$ and $\hat{R}_2(\theta,\varphi)$. 
In the four-dimensional Hilbert space spanned by the joint basis states of the two $2$-level atoms, the unitary transformation $\hat{R}_1(\theta,\varphi)$ effected by a Ramsey pulse on atom $1$ is
\begin{equation}
\label{eq:rams1}
\hat{R}_1(\theta,\varphi)=\hat{R}(\theta,\varphi)\otimes \hat{I},
\end{equation}
while a Ramsey pulse on atom $2$ is
\begin{equation}
\label{eq:rams2}
\hat{R}_2(\theta,\varphi)=\hat{I} \otimes \hat{R}(\theta,\varphi),
\end{equation}
where $\otimes$ denotes the tensor product operation.

The entangling operations, on the other hand, can be realized using the interactions between atoms and a cavity field governed by the two-atom Tavis-Cummings Hamiltonian in the limit of large detuning~\cite{PhysRevLett.85.2392,mseveritt09}. This produces an effective Hamiltonian in which the field is removed as a degree of freedom, eliminating atom-field entanglement, but allowing
virtual excitation of the field to pass excitations between atoms. This ensures that no quantum information is exchanged between the atoms and the cavity, so that the cavity merely mediates interactions between the atoms.
Each atom is effectively a two-state system, detuned from the cavity resonance by $\Delta$. Let $g$ denote the atom-cavity dipole coupling constant.
In the limit of large $\Delta$, the effective Hamiltonian is 
\begin{equation}
\label{eq:tchamilt}
\widetilde H_{\rm eff}  =  - \frac{{\hbar g^2 }}{\Delta }\left( {\sigma _{eg,eg}  + \sigma _{ge,ge}  + \sigma _{eg,ge}  + \sigma _{ge,eg} } \right).
\end{equation}
This yields a unitary transformation
\begin{equation}
\label{eq:dtc2}
\hat{T}_{\rm TC}(t)=\left[
\begin{array}{cccc}
 1 & 0 & 0 & 0 \\
 0 & e^{-i t \phi } \cos (t
   \phi ) & -i e^{-i t \phi
   } \sin (t \phi ) & 0 \\
 0 & -i e^{-i t \phi } \sin (t
   \phi ) & e^{-i t \phi }
   \cos (t \phi ) & 0 \\
 0 & 0 & 0 & 1
\end{array}
\right],
\end{equation}
where $\phi=g^2/\Delta (s^{-1})$ is the effective coupling constant and  we have used the notation $|gg\rangle \equiv [1,0,0,0]^T$, $|ge\rangle \equiv [0,1,0,0]^T$, $|eg\rangle \equiv [0,0,1,0]^T$ and $|ee\rangle \equiv [0,0,0,1]^T$ for the computational basis states.

To elucidate the process of deriving the superadditive measurement using these building blocks, it is instructive to first consider the Bell measurement briefly. The Bell measurement can be performed with a combination of the operations in Eqs.~\eqref{eq:rams1}, \eqref{eq:rams2} and \eqref{eq:dtc2}. This will yield a transformation which rotates each of the Bell states, into some permutation of the computational basis states up to global phases. 

This can be done in only four steps. The principle of the process is to take the entangled states to states which are as close as possible to the separable basis states with each step. 
The Tavis-Cummings operations are the available two-qubit operations for disentangling the Bell states. 
However, it turns out that the detuned Tavis-Cummings operation on its own cannot disentangle $|\Psi^\pm\rangle$. It is necessary to precede it with a Ramsey rotation which produces a relative phase shift of ${\pi}/{2}$ between $|01\rangle$ and $|10\rangle$. Previous work done have achieved the phase shifting effect using an extra atomic level $|i\rangle$. This is done by performing a Ramsey operation resonant with the $|e\rangle \leftrightarrow |i\rangle$ transition, before, and then after the Tavis-Cummings operation~\cite{PhysRevLett.85.2392, RaimondHaroche}. Another proposal suggests introducing a slight delay between the passage of the two atoms through the cavity~\cite{PhysRevA.77.023818}. Here we use another approach.

As a first step, we apply a Ramsey pulse to the atom $2$, i.e. $\hat{U}_1=\hat{R}_2\left(\pi,\frac{3\pi}{4}\right)$. This gives the following transformation of the Bell states:
\begin{eqnarray}
\label{eq:bellmstep1}
	\hat{U}_1|\Phi^\pm\rangle &=&  \left( |01\rangle\pm i|10\rangle\right)/{\sqrt{2}}  \nonumber\\
	\hat{U}_1|\Psi^\pm\rangle &=& \left( |00\rangle\pm i|11\rangle\right)/{\sqrt{2}}.
\end{eqnarray}

We then choose the second step as $\hat{U}_2=\hat{T}_{\rm TC}(\frac{3\pi}{4\phi})$ to have
\begin{eqnarray}
\label{eq:bellmstep2}
	\hat{U}_2\hat{U}_1|\Phi^+\rangle &=& \left(|00\rangle-i|11\rangle\right)/\sqrt{2}  \nonumber\\
	\hat{U}_2\hat{U}_1|\Psi^+\rangle &=&   |01\rangle  \nonumber\\
	\hat{U}_2\hat{U}_1|\Phi^-\rangle &=&  |10\rangle \nonumber\\
	\hat{U}_2\hat{U}_1|\Psi^-\rangle &=&  \left(|00\rangle+i|11\rangle\right)\sqrt{2}.
\end{eqnarray}
The combination of a preceding Ramsey operation on atom $2$ and the detuned Tavis-Cummings operation effectively carries out a transformation which disentangles the $|\Psi^+\rangle$ and $|\Phi^-\rangle$ states. 

We now proceed to the third step which is a Ramsey operation $\hat{U}_3=\hat{R}_2\left(\pi,0\right)$. The effect of this operation is simply to interchange  $|00\rangle$ with $|01\rangle$, and $|11\rangle$ with $|10\rangle$ simultaneously. 
\begin{eqnarray}
\label{eq:bellmstep3a}
	\hat{U}_2\hat{U}_1|\Phi^+\rangle &=& \left(|01\rangle-i|10\rangle\right)/\sqrt{2}  \nonumber\\
	\hat{U}_2\hat{U}_1|\Psi^+\rangle &=&   |00\rangle  \nonumber\\
\label{eq:bellmstep3b}
	\hat{U}_2\hat{U}_1|\Phi^-\rangle &=&  |11\rangle \nonumber\\
	\hat{U}_2\hat{U}_1|\Psi^-\rangle &=&  \left(|01\rangle+i|10\rangle\right)\sqrt{2}.
\end{eqnarray}
The resulting entangled states can now be disentangled with a Tavis-Cummings operation in a final step before detection.
If we use  $\hat{U}_4=\hat{T}_d(\frac{3\pi}{4\phi})$, then we will have:
\begin{eqnarray}
\label{eq:bellmstep4}
	\hat{U}_4\hat{U}_3\hat{U}_2\hat{U}_1|\Phi^+\rangle &=&  |01\rangle  \nonumber\\
	\hat{U}_4\hat{U}_3\hat{U}_2\hat{U}_1|\Psi^+\rangle &=&  |00\rangle  \nonumber\\
	\hat{U}_4\hat{U}_3\hat{U}_2\hat{U}_1|\Phi^-\rangle &=&  |11\rangle \nonumber\\
	\hat{U}_4\hat{U}_3\hat{U}_2\hat{U}_1|\Psi^-\rangle &=&  |10\rangle.
\end{eqnarray}
Detection results $|00\rangle$, $|01\rangle$, $|10\rangle$ and $|11\rangle$ would indicate that the inputs were the Bell states  $|\Psi^+\rangle$, $|\Phi^+\rangle$, $|\Psi^-\rangle$ and $|\Phi^-\rangle$ respectively. This realisation is similar in its construction to the realisation on atoms considered in~\cite{PhysRevA.62.052311}.

We now apply a similar method for the superadditive measurement. It is worth noting that since the outcome corresponding to $|A\rangle$ should never occur, it would be sufficient to make a unitary transformation $\hat{U}_{\rm sa}'$ which takes two of the three states $|\Pi_{00}\rangle$, $|\Pi_{11}\rangle$ and $|\Pi_{22}\rangle$ uniquely into two of the four computational basis states in the four-dimensional Hilbert space, say $|\Pi_{22}\rangle\rightarrow|00\rangle$ and $|\Pi_{11}\rangle\rightarrow|01\rangle$; and the remaining measurement basis state, say $|\Pi_{00}\rangle$, and $|A\rangle$ each into superpositions of the two other computational basis states, say $|10\rangle$ and $|10\rangle$. This may somewhat simplify the experimental realization, and we will in fact make use of it.

Obtaining a realization of $\hat{U}_{\rm sa}$ involves finding a sequence of operations which transforms $\hat{U}^\dagger_{\rm sa}$ into a matrix of the form $P_\pi D$, where $P_\pi$ is a permutation matrix and $D$ is a diagonal matrix. This could be a sequence of unitary operations coupling two basis states at a time~\cite{PhysRevLett.73.58}. As with most physical settings, not all pairwise coupling operations are available in our case. This is because we are restricted to operations $\hat{T}_{\rm TC}(t)$, which couple the pair of basis states $|01\rangle$ and $|10\rangle$, and single qubit operations $\hat{R}_1(\theta,\varphi)$ and $\hat{R}_2(\theta,\varphi)$, which each couple two pairs of basis states at the same time. Our strategy for obtaining a realisation in terms of these is as follows. Since the operation $\hat{T}_{\rm TC}$ couples basis states $|01\rangle$ and $|10\rangle$, it is natural to first use a Tavis-Cummings interaction to disentangle these components of the measurement states in Eq.~\eqref{eq:mbas14}. 
In order to do this, it turns out that we need to precede the Tavis-Cummings interaction by two Ramsey pulses.
This first pulse sequence then takes the states $|A\rangle=|\Psi^-\rangle$ and $|\Psi^+\rangle$ into the disentangled states $|01\rangle$ and $|10\rangle$. Next, in order to use a Tavis-Cummings interaction to disentangle the $|00\rangle$ and $|11\rangle$ components, we need to first change $|0\rangle$ into $|1\rangle$ and vice versa for one of the atoms, which one does not matter, using Ramsey pulses. It turns out that at the end of this process, which thus comprises two Tavis-Cummings interactions and a number of Ramsey pulses, $|A\rangle$ and the measurement basis state $|\Pi_{00}\rangle$ are both mapped to superpositions of $|00\rangle$ and $|10\rangle$, and Ramsey pulses would be needed in order to map these superpositions to $|00\rangle$ and $|10\rangle$. As remarked above, these last Ramsey rotations are not necessarily required.

This leads us to a realization in seven steps.
The first step is a Ramsey rotation on the atom 2, $\hat{U'_1}=\hat{R}_2\left(\pi,\pi\right)$. The second step is another Ramsey rotation on atom 2, $\hat{U'_2}=\hat{R}_2\left(\pi ,{3 \pi}/{4}\right)$. For the third step, we pass the two atoms simultaneously through the first detuned cavity with the effective interaction time $t_1={3\pi}/{(4\phi)}$ giving a detuned Tavis-Cummings interaction described by $\hat{U'_3}=\hat{T}_{\rm TC}(t_1)$. Step four is another Ramsey pulse applied to atom 2 defined as  $\hat{U'_4}=\hat{R}_2\left(\pi ,{\pi}/{2}\right)$. The fifth step is a second detuned Tavis-Cummings type interaction, $\hat{U'_5}=\hat{T}_d(t_2)$, of duration $t_2=\frac{\gamma}{2\phi}$. The sixth and seventh steps effectively rotate $|A'\rangle$ into a superposition. In fact, this takes place if we choose $\hat{U'_6}=\hat{R}_2\left(\pi ,\left(\gamma -{\pi}/{2}\right)/4\right)$ as the sixth step  and $\hat{U'_7}=\hat{R}_2\left({\pi}/{2},0\right)$ as the seventh step.

These steps lead us to the effective unitary operation
\begin{equation}
\label{ed:resadd82}
\hat{U}_{\rm sa}'=\frac{e^{-i\gamma/4}}{2}\left[
\begin{array}{cccc}
 -\sqrt{2}e^{-i \gamma/2} C &
   -e^{i\gamma/2}&
    e^{i\gamma/2} &
   \sqrt{2}e^{-i \gamma/2} S\\
 -\sqrt{2}e^{-i \gamma/2} C &
   e^{i\gamma/2}&
   - e^{i\gamma/2} &
   \sqrt{2}e^{- i \gamma/2} S \\
 - \sqrt{2}S&
 1 &
1 &
   -\sqrt{2}C\\
 \sqrt{2}S&
1 &
1 &
    \sqrt{2}C
\end{array}
\right],
\end{equation}
where $S= \sin(\gamma/2)$ and $C=\cos(\gamma/2)$. 
To elucidate the assignment of measurement results, we can also use the alternative form
\begin{eqnarray}
\label{ed:resadd9}
\hat{U}_{\rm sa}'  \equiv &{\tfrac{1}{{\sqrt 2 }}}\left( {\left| {00} \right\rangle  + \left| {10} \right\rangle } \right)\left\langle {\Pi _{00} } \right| + {\tfrac{1}{{\sqrt 2 }}}\left( {\left| {00} \right\rangle  - \left| {10} \right\rangle } \right)\langle A| \nonumber\\
&+ \left| {10} \right\rangle \left\langle {\Pi _{22} } \right| + \left| {11} \right\rangle \langle \Pi _{11} |.
\end{eqnarray}
This makes it clear that $\hat{U}_{\rm sa}'$ ideally yields the same value as $\hat{U}_{\rm sa}$ for the mutual information, but is slightly different from $\hat{U}_{\rm sa}$, since final detection of $|00\rangle$ and $|10\rangle$ both correspond to $|\Pi_{00}\rangle$. Recall that all three signal states are orthogonal to $|A\rangle$. 
When $\hat{U}_{\rm sa}$ is used, the final measurement outcome corresponding to $|A\rangle$ should never occur. When experimental imperfections are included, the mutual information and consequently SQCG may be different for $\hat{U}_{\rm sa}$ and $\hat{U}_{\rm sa}'$. This will be made clear shortly.

After performing the seven steps, a subsequent detection in the computational basis will complete the measurement. 
The mutual information is given by
\begin{eqnarray}
I_2&=&I(X:Y)  \label{eq:mutinfo}\\
&=&\sum\limits_j {P(x_j)\sum\limits_k {P(y_k |x_j)\log _2 } } \left[ {\frac{{P(y_k |x_j)}}{{\sum\limits_m {P(x_m)P(y_k |x_m)}}}}\right],
\nonumber
 \end{eqnarray}
where $X$ and $Y$ denote the sender and receiver respectively, and $x_j$ and $y_k$ denote the letter states that were transmitted and received respectively, $k,j,m=1,2,3,...$.

\begin{figure*}
\centerline{\includegraphics[width=0.8\textwidth]{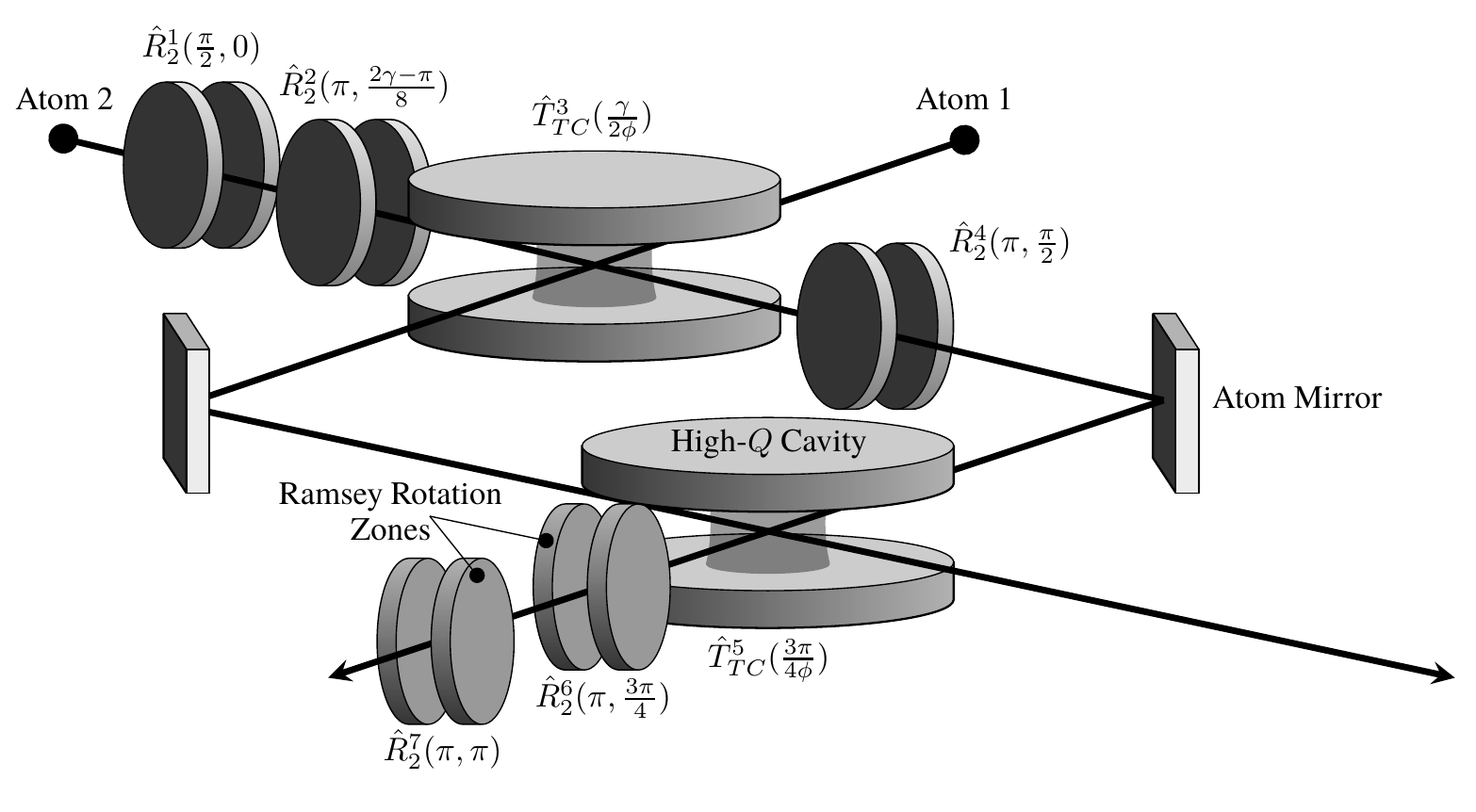}}
\caption{Schematic diagram of proposed cavity QED implementation of the POM for superadditive decoding. This figure shows the scheme for the unitary operation performed before the final projective measurement by field-ionisation detectors. Atom mirrors have been implemented using electric, magnetic and light-induced forces (see \cite{PhysRevLett.97.033002}, \cite{0953-4075-39-18-002} and references therein). If required, it could be realised in this scheme also by using an additional Tavis-Cummings operation to swap the state from one atom to that of a fresh one travelling perpendicularly. The superscripts on the operators indicate the order of their application to the two atom state. These operations are defined in Equations \eqref{eq:rams1}, \eqref{eq:rams2}, and \eqref{eq:dtc2}.}
\label{fig:cqedsetup3d}
\end{figure*}

The channel matrix resulting from applying the derived pulse sequence and a subsequent projective measurement to the input state $\hat{\rho}_x=|\psi_x\rangle\langle \psi_x|$ are
\begin{equation}%
\label{eq:chmat}
P\left({y|x}\right)={\rm Tr}(\hat{\Pi'}_{y}\hat{\rho}_{x}),
\end{equation}
where $\hat{\Pi'}_{y}=|{\Pi'}_{yy}\rangle\langle{\Pi'}_{yy}|$. Substituting the resulting channel matrix elements into Eq.~\eqref{eq:mutinfo}, and using prior probabilities 
\begin{equation}
\label{eq:proir}
p\left( {x_j}\right)=\frac{1}{3},~~~~(j=1,2,3)
\end{equation}
gives SQCG of  $I_2/2-C_1=0.0391$.
A schematic diagram outlining the derived implementation is shown in Fig.~\ref{fig:cqedsetup3d}. 

\subsection{Optimality}
The question of determining the optimality of a given realization of a POM using certain building blocks in a physical setting is non-trivial. For our realization of the superadditive measurement, we check for the optimality of our proposed realization in terms of the total number of steps. We also try to exclude the more experimentally challenging steps as much as possible. The Tavis-Cummings interaction is clearly more difficult to realize than the Ramsey operations because it involves a synchronous passage of two atoms through a high-Q cavity, which is more experimentally challenging than applying a Ramsey pulse to a single atom. 

We supply a short proof by contradiction that at least two detuned Tavis-Cummings interactions is required to realize the superadditive decoding. 
Using the canonical Cartan decomposition of a two-qubit unitary operator ${\rm{U}} \in {\rm{SU(4)}}$~\cite{PhysRevA.67.042313}, we realize that if a single detuned Tavis-Cummings interaction could be used to implement $\hat{U}_{\rm sa}$ (or $\hat{U}_{\rm sa}'$), then there would exist $w'_1$, $w_2$, $v'_1$, $v_2$, $W'_1$, $W_2$, $V'_1$, and $V_2$ $\in\rm{SU(2)}$ such that
\begin{equation}
\label{eq:kak}
\hat{U}_{\rm sa} =(w'_1\otimes w_2) T_d(\phi) (v'_1\otimes v_2)
\end{equation}
and
 \begin{equation}
\label{eq:kak2}
T_d(\phi)=(W'_1\otimes W_2) \hat{U}_{\rm sa} (V'_1\otimes V_2).
\end{equation}
Equation~\eqref{eq:kak2} is a system of $16$ equations. It is easily verified that this system of equations has no solution. This concludes the proof and gives evidence of the optimality of our proposed scheme with respect to the number of Tavis-Cummings interactions needed.
Jaynes-Cummings interactions through sequential passage of the atoms through the cavity can also be used for entangling interactions between atoms. This has proved suitable for preparing specific entangled two-atom states~\cite{PhysRevLett.79.1,PhysRevA.77.023818,PhysRevLett.85.2392}. However, a main disadvantage of using the Jaynes-Cummings interactions is the leakage of atomic excitation into cavity field modes having more than one excitation, since the field has an infinite number of levels besides $\{|0\rangle, |1\rangle\}$.

\section{Experimental imperfections}

Both the IDP measurement and the measurement to demonstrate quantum superadditivity will be affected by experimental imperfections. In particular, when errors are present, error-free or unambiguous state discrimination in general becomes impossible, and we should aim for a maximum confidence measurement strategy instead~\cite{PhysRevLett.96.070401, PhysRevA.77.012113, PhysRevLett.97.193601}. As for the measurement that demonstrates superadditivity, it is natural to ask how robust the superadditive quantum coding gain is with respect to imperfections. We will now discuss this.

Experimental imperfections that could adversely affect the overall quality of the realizations of  the superadditive measurement, and the SQCG, include initial state preparation fidelity, Ramsey operation fidelity, Tavis-Cummings operation fidelity and detection efficiency. The initial state preparation fidelity would depend largely on the fidelity of the Ramsey operations since they are used to carry out these preparations. In turn, the fidelity of the Ramsey operations depend on the accuracy to which the parameters $\theta$ and $\varphi$ can be set. 

Let us consider how the delay between the atoms affect the results and ultimately the SQCG. Zheng and Guo~\cite{PhysRevLett.85.2392} have estimated the effect of such a delay on the preparation of an EPR pair of the form 
\begin{equation}
\label{eq:epr}
|\Psi_{EPR}\rangle = \frac{1}{\sqrt{2}}\left(|e,g\rangle+ i |g,e\rangle\right),
\end{equation}
which can be prepared by a single Tavis-Cummings operation.
This was done by considering  a delay of $t_d=0.01t$ between the atoms, where $t$ is the time each atom spends in the cavity. In this situation, a fidelity of $0.99$ was estimated.

\begin{figure}
\centerline{\includegraphics[width=0.35\textwidth]{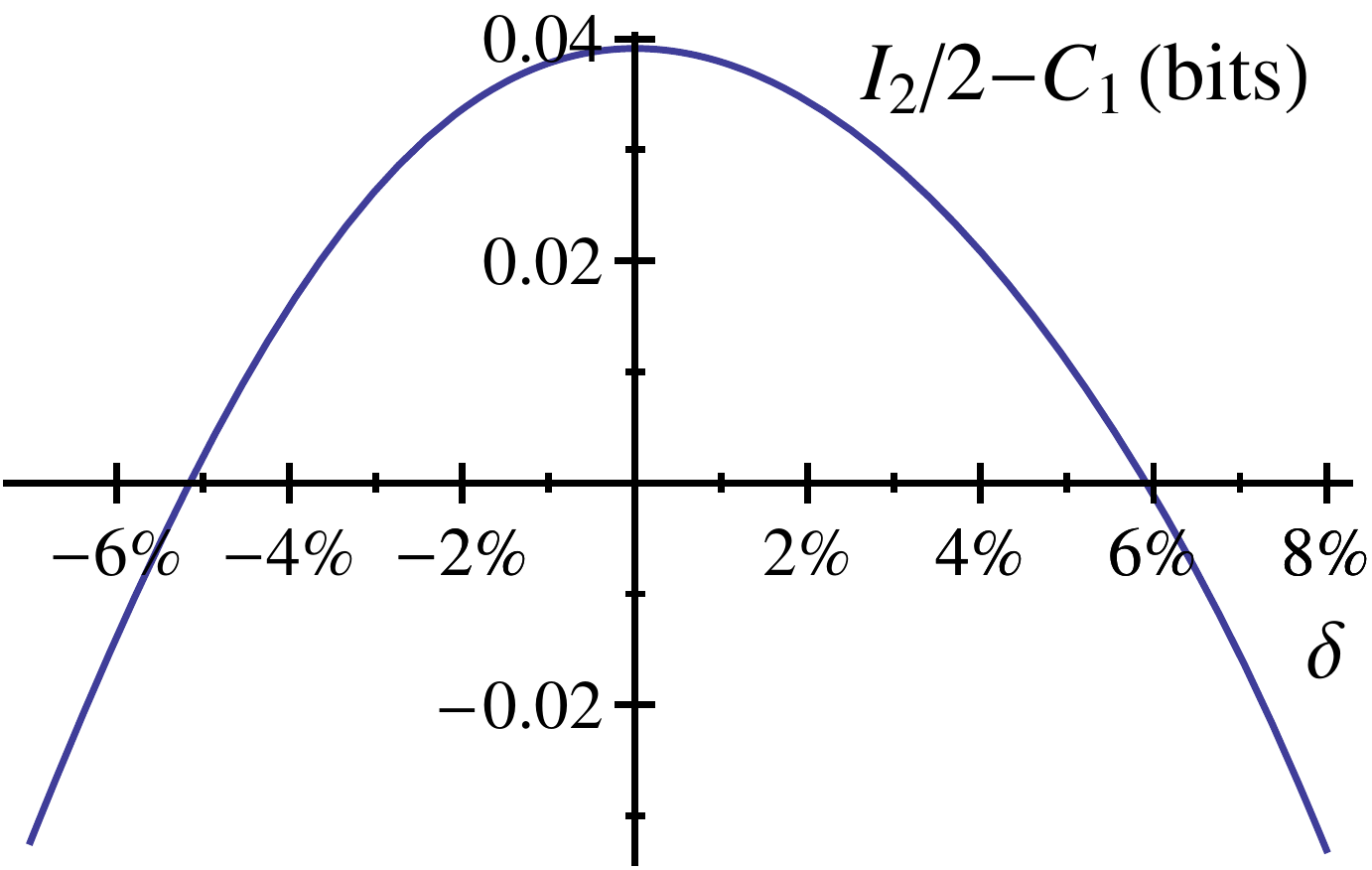}}
\caption{Plot of the superadditive coding gain as a function of $\delta$, where $\delta$  is the delay $t_d (s)$ as a percentage of the time $t_2$ (s) spent by both atoms in the second cavity. The same delay is applied in both cavities.}
\label{fig:delayatom}
\end{figure}

Applying the same idea, we realize that such a delay yields an imperfect Tavis-Cummings operation which affects the coding gain. In Fig.~\ref{fig:delayatom}, we plot the superadditive coding gain as a function of $\delta$, where $\delta$  is the delay as a percentage of the longest cavity interaction time in the sequence, that is, $t_2$ (s) spent in cavity 2. A delay up to $5\%$ of the longest cavity interaction time in the sequence, which occurs in the second cavity interaction, still gives an SQCG of $0.011 $ bits. 

In the photonic realization~\cite{PhysRevA.69.052329}, the detection efficiency $\eta$, which is the photon count probability, does not degrade the result on its own. This is because the SQCG is calculated using a normalised channel matrix. However, when combined with dark counts which arise from background radiation as well as from carriers generated in a detector even when no photons are incident, the SQCG is degraded since this effectively results in a finite probability of misidentification of states. 

In a cavity QED realization, the detection efficiency could even be more of a problem if it depends on the atomic states, for example. Even if the detection efficiency were independent of the atomic state, state misidentification is a usual problem in detection. Consider a detection to determine whether a two level atom is in a state $|0\rangle$ or $|1\rangle$. In the perfect case, the detection would be an ideal von Neumann measurement which can be described by the two projectors
\begin{equation}%
\hat{P}_0=|0\rangle\langle 0 |, ~~\hat{P}_1=|1\rangle\langle 1 |.
\label{eq:detvonNeum}
\end{equation}
A non-ideal detector, however, might record the wrong state with some probability. This is the case in atomic state detection schemes where projective measurements are carried out using field-ionisation detectors, in which the ionisation energy of the atoms serves as an indicator of the state. This means that for a two-level atom in the state $|0\rangle$, the measurement will give the result $1$ with probability $p$ and the result $0$ with probability $1-p$. In a realistic experimental scenario, the probability of misidentification might not be symmetric. For instance, it might be more likely to misidentify the atomic state $|1\rangle$ as $|0\rangle$ than conversely. Let us denote the probability of misidentifying the $|0\rangle$ and $|1\rangle$ states as $p(0|1)\equiv p$ and $p(1|0)\equiv q$ respectively. Introducing these errors, the effective measurement is a POM with elements defined as the operators
\begin{eqnarray}%
\hat{\pi}_0=(1-p)\hat{P}_0+q\hat{P}_1 \nonumber \\
\hat{\pi}_1=(1-q)\hat{P}_1+p\hat{P}_0.
\label{eq:nonidealpom}
\end{eqnarray}

\begin{figure}[floatfix]
     \flushleft
\includegraphics[width=0.48\textwidth]{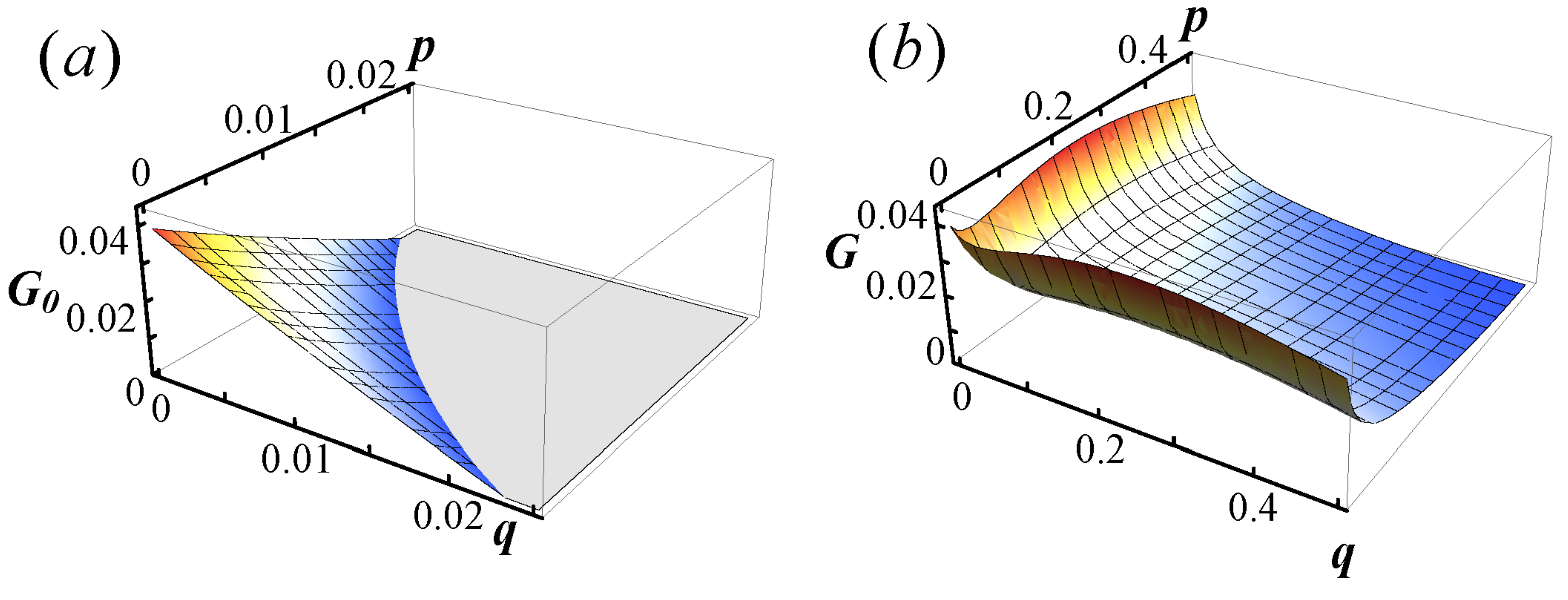}
\caption{Effects of detection errors when $\hat{U}_{\rm sa}$ is used to realize the superadditive measurement. (a) Plot of the difference $G_0$ (bits) between the length-two coding mutual information and the ideal single channel capacity, $G_0=I_2/2(p,q) - C_1(p=0,q=0)$. (b) Plot of the actual superadditive coding gain $G_0$ (bits), $G=I_2/2(p,q) - C_1(p,q)$. $G_0$ and $G$ are plotted as functions of the probabilities, $p$ and $q$, of misidentifying states $|0\rangle$ and $|1\rangle$ of a two-level atom respectively.}
    \label{fig:sqcgimp}
\end{figure}

To incorporate this into the calculations of the SQCG, we first calculate the resulting single channel capacity $C_1$ and mutual information for length-two coding $I_2$. This is used to obtain the SQCG plotted in Fig.~\ref{fig:sqcgimp} as a function of the probabilities of misidentification $p$ and $q$ when $\hat{U}_{\rm sa}$ is realized exactly. The affected channel matrix is given as
\begin{equation}%
\label{eq:chmats}
P\left({y|x}\right)={\rm Tr}(\hat{U}_{\rm sa}\hat{\rho_x}\hat{U}_{\rm sa} ^\dag \hat{M}_y).
\end{equation}
Here $x$ and $y$ label the matrix elements. $\hat{M}_1=\hat{\pi}_0\otimes\hat{\pi}_0$, $\hat{M}_2=\hat{\pi}_0\otimes\hat{\pi}_1$, $\hat{M}_3=\hat{\pi}_1\otimes\hat{\pi}_0$, and $\hat{M}_4=\hat{\pi}_1\otimes\hat{\pi}_1$ are the elements of the POM describing the imperfect projective measurement in the computational basis.
The resulting SQCG plot shows that even with rather high probabilities of misidentification, a reasonable amount of superadditive quantum coding gain can still be accessed. We observe a symmetric trade-off effect between probabilities $p$ and $q$ in Fig.~\ref{fig:sqcgimp}. 

\begin{figure}[floatfix]
\includegraphics[width=0.48\textwidth]{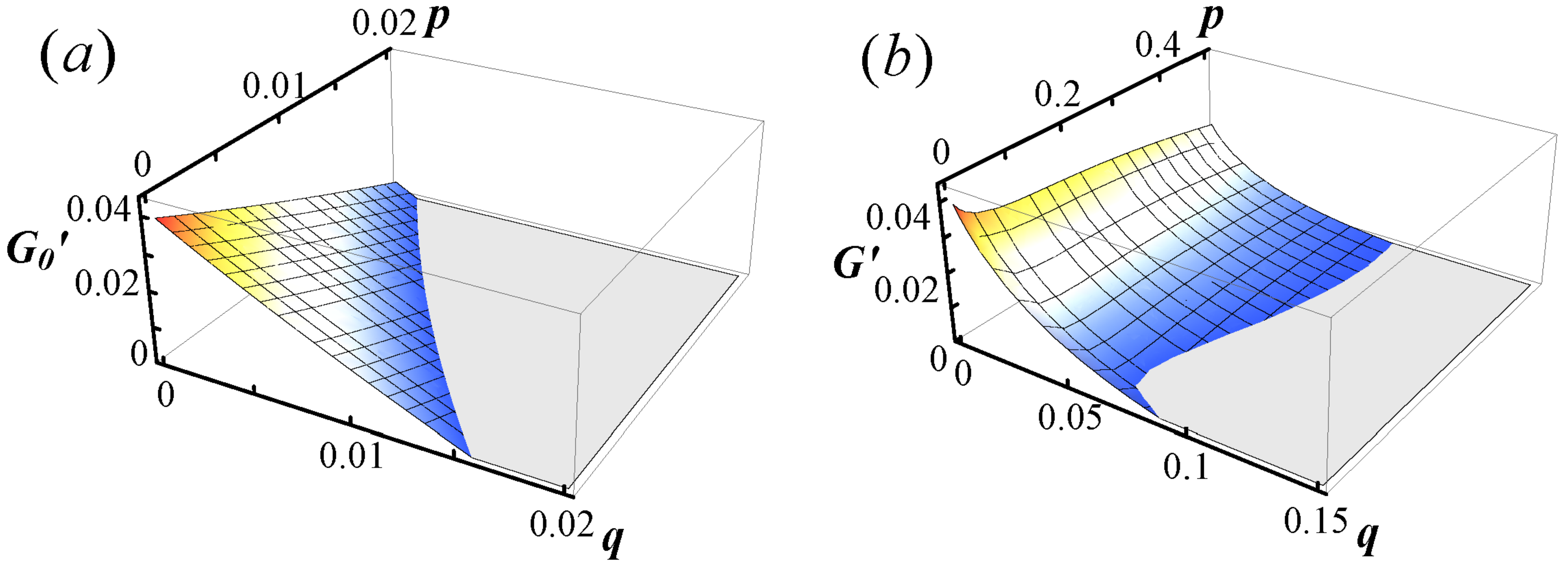} \label{fig:sqcgimp3b}
\caption{Effects of detection errors when $\hat{U}_{\rm sa}'$ is used to implement the superadditive measurement. (a) Plot of the difference $G_0'$(bits) between the length-two coding mutual information and the ideal single channel capacity, $G_0'=I_2'/2(p,q) - C_1(p=0,q=0)$. (b) Plot of the actual superadditive coding gain $G'$ (bits) for our proposed scheme, where $G'=I_2'/2(p,q) - C_1(p,q)$. $G_0'$ and $G'$ are plotted as functions of the probabilities, $p$ and $q$, of misidentifying states $|0\rangle$ and $|1\rangle$ of a two-level atom respectively.}
    \label{fig:sqcgimp3}
\end{figure}

Let us now use the proposed realization which effects the unitary transformation $\hat{U}_{\rm sa}'$. The channel matrix elements then become
\begin{equation}%
\label{eq:chmats2}
P\left({y|x}\right)={\rm Tr}(\hat{U}_{\rm sa}'\hat{\rho_x}\hat{U}_{\rm sa}'^\dag \hat{M}_y).
\end{equation}
The corresponding values of mutual information for the double channel and the SQCG are plotted in Fig.~\ref{fig:sqcgimp3} as functions of the probabilities of misidentification $p$ and $q$.

As shown in Fig.~\ref{fig:sqcgimp3}(b), we observe reasonable amounts of SQCG even with rather high levels of detection errors. Since, in our proposed scheme $\hat{U}_{\rm sa}'$, the SQCG favours combinations of higher values of $p$ with lower values $q$, the physical states representing $|0\rangle$ and $|1\rangle$ may need to be chosen to ensure that $p>q$ if there is considerable difference between $p$ and $q$. 

Finally, another experimental consideration in the cavity QED realizations outlined above could be the symmetry of the Ramsey operations, that is, depending on the particular experimental setup, whether or not there is an advantage of performing the Ramsey rotations on only one atom over distributing them as much as possible between both atoms.  

\section{Conclusion}
In conclusion, we have proposed explicit schemes for experimental realization, using cavity QED, of two generalized quantum measurement strategies. These were unambiguous discrimination of two non-orthogonal quantum states, the so-called IDP measurement, and the measurement to demonstrate superadditive quantum coding using a ternary quantum alphabet. We would like to note that realizations of the minimum-error measurements to distinguish between the trine states in Eq.~\eqref{eq:trine3}~\cite{helstrom}  and between mirror-symmetric states~\cite{PhysRevA.65.052308} would be very similar to the IDP measurement that we have outlined, and also that similar methods can be used to implement any generalized quantum measurement using cavity QED.

 Our results show that these realizations are feasible using currently available cavity QED technologies. Using a simple proof we have confirmed the optimality of the realization of the measurement that demonstrates quantum superadditivity in terms of cavity usage. We have also shown how the superadditive quantum coding gain is affected by imperfect detection of the basis states, and that even with rather high levels of such experimental imperfections, a reasonable amount of superadditivity can be seen. We have not addressed the fact that in the presence of experimental imperfections, the measurement that one should attempt  to implement in order to demonstrate maximum coding gain might change. It is thus conceivable that even with experimental errors, it may be possible to see a somewhat larger quantum coding gain than our estimates indicate. In other words, our estimates are lower bounds on the superadditive quantum coding gain, given the assumed level of errors in the implementation. An example where the optimal quantum measurement changes in the presence of experimental imperfections is when comparing two coherent states~\cite{PhysRevA.79.023808}.
 
The fact that atoms can interact strongly via cavity fields makes it possible to investigate implementation of superadditive coding with longer code words using cavity QED-type systems. It is also interesting to further study realizations of other generalized quantum measurements which are difficult to realize using linear optics.

\section{Acknowledgements}
AD is funded by a Scottish Universities Physics Alliance (SUPA) scholarship, ME acknowledges support from the Japanese Society for the Promotion of Science, and VMK is funded by a UK Royal Society University Research fellowship.

%


\end{document}